\documentclass[twocolumn]{aastex63}

\received{2020 September 10}
\accepted{2020 November 20}
\shorttitle{Refined Telluric Absorption Correction}
\shortauthors{Kimeswenger et al.}

\begin{document}

\title{Refined Telluric Absorption Correction for Low-Resolution 
Ground-Based Spectroscopy:\\ Resolution and Radial Velocity Effects in the O$_2$ A-Band for Exoplanets and \ion{K}{1} Emission Lines}

\correspondingauthor{Stefan Kimeswenger}
\email{Stefan.Kimeswenger@uibk.ac.at}

\author[0000-0003-2379-0474]{Stefan Kimeswenger}
\affiliation{
Institut f{\"u}r Astro- und Teilchenphysik, Universit{\"a}t Innsbruck, Technikerstr.~25/8, 6020 Innsbruck, Austria}
\affiliation{Instituto de Astronom{\'i}a, Universidad Cat{\'o}lica del Norte, Av. Angamos 0610, Antofagasta, Chile}

\author{Manuel Rainer}
\affiliation{
Institut f{\"u}r Astro- und Teilchenphysik, Universit{\"a}t Innsbruck, Technikerstr.~25/8, 6020 Innsbruck, Austria}

\author[0000-0001-5263-9998]{Norbert Przybilla}
\affiliation{
Institut f{\"u}r Astro- und Teilchenphysik, Universit{\"a}t Innsbruck, Technikerstr.~25/8, 6020 Innsbruck, Austria}

\author[0000-0003-3557-7689]{Wolfgang Kausch}
\affiliation{
Institut f{\"u}r Astro- und Teilchenphysik, Universit{\"a}t Innsbruck, Technikerstr.~25/8, 6020 Innsbruck, Austria}



\begin{abstract}
Telluric correction of spectroscopic observations is either performed via standard stars that are observed close in time and airmass along with the science target, or recently growing in importance, by theoretical telluric absorption modeling. Both approaches work fine when the telluric lines are resolved, i.e.~at spectral resolving power larger than about 10\,000, and it is sufficient to facilitate the detection of spectral features at lower resolution. However, a meaningful quantitative analysis 
requires also a reliable recovery of line strengths. Here, we show for the  Fraunhofer A-band of molecular O$_2$ that the standard telluric correction approach fails in this at lower spectral resolutions, as an example for the general problem. Doppler-shift dependent errors of the restored flux may arise, which can amount to more than 50\% in extreme cases, depending on the line shapes of the target spectral features. Two applications are discussed: the recovery of the O$_2$-band in the reflected light of an Earth analog 
atmosphere, as facilitated potentially in the future using an orbiting starshade and a ground-based extremely large telescope; and the recovery of the intrinsic ratio of the K\,{\sc i} lines in the post-nova V4332 Sgr tracing the optical depth of the emitting region, to exemplify the relevance using present-day instrumentation. We show how one should derive correction functions for the compensation of the error in dependence of radial velocity shift, spectral resolution and target line-profile function by use of high resolution atmospheric transmission modeling, which has to be solved for the individual case. 
\end{abstract}


\section{Introduction} \label{sec:intro}

Ground-based spectroscopy can be strongly affected by absorption 
caused by molecules in the Earth's atmosphere. Several observational approaches have been suggested  to correct target spectra for telluric absorption lines. Proposed were supplementary observations of either fast-rotating early-type stars \citep{Vidal-Madjaretal86} or G2\,V stars \citep{Maiolinoetal96} as so-called telluric standard stars (TSS). The derived telluric correction spectrum may be refined further in a second stage by correcting also for the stellar spectral features \citep[e.g.][]{Vaccaetal03,Sameshimaetal18}, or for the solar spectrum. Ideally, the target spectrum is recovered as it would appear if the instrument was located above the atmosphere. 

The correction procedure is always the following. The TSS with known intrinsic spectral flux $I_{\rm TSS}(\lambda)$ is observed just before and/or after the target. It facilitates the atmospheric transmission $T(\lambda,a)$ as a function  of wavelength and airmass $a$ to be derived
\begin{equation}\label{equ_1}
    T(\lambda,a) = \frac{F_{\rm TSS}(\lambda,a)}{I_{\rm TSS}(\lambda)}\,,
\end{equation}
where $F_{\rm TSS}$ is the observed flux of the TSS corrected for instrumental effects. 
The target's intrinsic flux $I_{\rm target}(\lambda)$ is then reconstructed from the observed target flux $F_{\rm target}(\lambda,a)$ as 
\begin{equation}\label{equ_2}
    I_{\rm target}(\lambda) = \frac{F_{\rm target}(\lambda,a)}{T(\lambda,a)}\,.
\end{equation}
Ideally, the TSS observations are taken at similar airmass $a$ as the science target to eliminate the dependency on that parameter. 

An alternative approach was presented with theoretical modeling of the atmosphere 
by FASCODE/LBLRTM \citep{Seifahrtetal10}, TAPAS \citep{Bertaux2014},  TelFit  \citep{Gullikson2014} or molecfit
\citep{Smette2015,Kausch2015}. Here, the function $T(\lambda,a)$ is calculated with theoretical radiative transfer models of the Earth's atmosphere, using ambient condition measurements and atmospheric profiles provided by international weather forecast centers. While TAPAS does not perform fitting, the other approaches can adapt the atmosphere profiles by optimizing the fit of telluric absorption features in the target observations directly. The different telluric correction approaches were recently compared by \citet{Ulmer-Moll19}, who concluded that {\sc i}) generally the modeling approach provides better corrections than the observational 
one and that {\sc ii}) mostly in their near-infrared study molecfit (which is employed in the present study) superseded the other methods. 

As long as the atmospheric absorption lines are resolved, the corrections can be obtained at high quality, like e.g.~for the prominent Fraunhofer A-band formed by the molecular O$_2$ of our Earth's atmosphere \citep{A_band}, on which we will concentrate in the following. This wavelength range has attracted limited interest in astrophysical research for a long time. However, it has come into focus recently, as observations in the A-band are highly promising for tracing O$_2$ in the atmosphere of a terrestrial exoplanet, which could potentially act as a biomarker if the planet is located in the star's habitable zone. This is nicely discussed in the study of very high resolution Doppler-shift techniques by \citet{LopezMorales2019} that may be employed for investigating transmission spectroscopy of Earth analogs during transits using ground-based extremely large telescopes (ELTs). A second criterion for achieving high-quality telluric correction requires the absence of strong variations of the target flux at wavelengths blocked by the telluric features \citep{Christiaens2018}. 

However, all the previously mentioned methods face complications, once the telluric lines are not resolved sufficiently. An important motivation for this case comes from photon-starved observations of exoplanets that are complementary to transmission spectroscopy during transits: spectroscopy of the reflected light of directly-imaged exoplanets, which is not restricted to edge-on viewing geometries.
These observations may become feasible for terrestrial exoplanets in the habitable zones of nearby stars with an orbiting starshade + ground-based ELTs \citep{Mather2019}. At an estimated brightness of an Earth analog exoplanet (henceforth called an exo-Earth) of $V\sim30\fm0$, this will be feasible only at low spectral resolution even with an ELT. The disentangling of the exo-Earth and terrestrial O$_2$ signatures will be difficult. However, there are differences in line depths and shapes compared to the single line-of-sight through the terrestrial atmosphere. They are introduced by a number of factors: integration of the spectrum over all contributions from the illuminated disk of the exo-Earth, 
different oxygen concentrations, multiple scattering, cloud coverage in the exo-Earth etc., and radial velocity induced wavelengths shifts.
Sophisticated telluric correction will be required for a meaningful interpretation of the observations.

In this paper we therefore systematically study the effects of the spectral resolving power and in addition of the radial velocity of the target on the quality of the telluric correction. The question is how well target features can be disentangled from absorption arising from the Earth's atmosphere. 
We first describe the methods employed in our study (Sect.~\ref{sec:methods}). Then, applications and results are discussed in Sect.~\ref{sec:applications}. The focus is on simulated data of observations of molecular O$_2$ in an Earth analog as observed within the A-band wavelength range. As a proof-of-concept we test our methods on real data taken at different spectral resolution for the recovery of the neutral potassium resonance lines which fall into the A-band wavelength region. Finally, conclusions are drawn  and an outlook is given in Sect.~\ref{sec:conclusions}. We want to note that while we concentrate on the A-band here, the results can easily be generalized to all spectral features that overlap with telluric molecular bands.

\section{Methods}\label{sec:methods}

\begin{figure}[t!]
\medskip
    \centering
    \includegraphics[width=85mm]{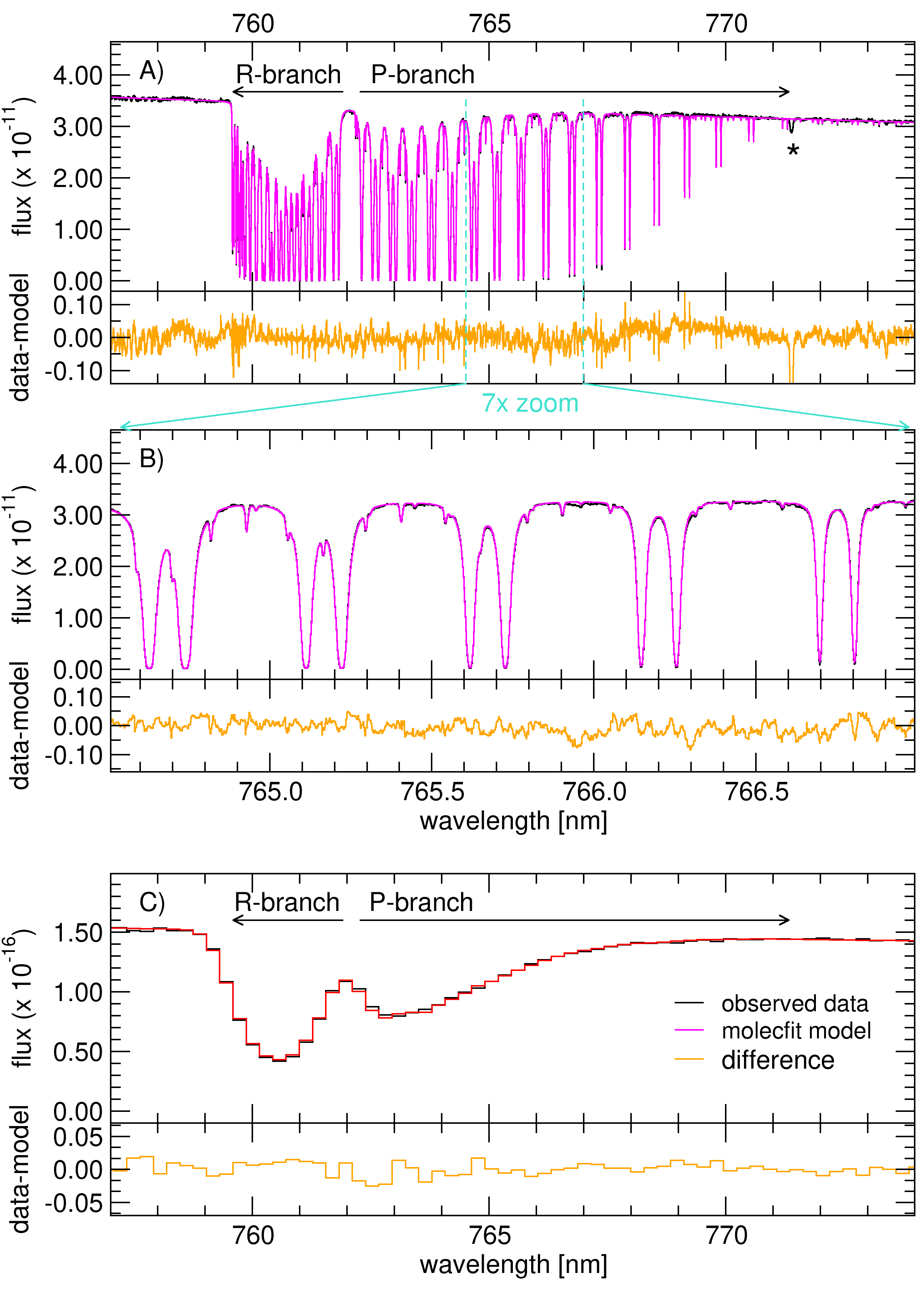}
    \caption{The telluric Fraunhofer A-band caused by molecular O$_2$ absorption. All panels show in the same way the observed data, a theoretical absorption spectrum obtained by molecfit and the difference: A) An ESO VLT UVES spectrum of Sirius; the asterisk ({\tt *}) marks a stellar spectral line not included in our model. B) The same spectrum as above, but zoomed to the region where the lines change from fully saturated to unsaturated absorption.  C) ESO VLT FORS2 spectrum of the white dwarf Feige\,110. All spectra are calibrated in SI units of W m$^{-2}$ nm$^{-1}$.}
    \label{fig:01}
\end{figure}

In a first step, we test our telluric correction methodology on targets with only continuous spectral flux within the Fraunhofer A-band in order to demonstrate its fidelity. 
The A-band is known to be very stable since the O$_2$ fraction is widely constant with time (about 20.9 per cent at sea level). The R-branch ($759.60<\lambda<761.90\,$nm)\footnote{The R-branch wavelength range is sharply defined by the physics of the bandhead.} of the A-band shows very dense regions of nearly fully absorbing lines.
The P-branch  ($762.10<\lambda<771.33\,$nm)\footnote{The P-branch wavelength range in the present case is defined by us as the region where absorption in the line cores is at least 2\% of the flux. The P-branch in fact extends in some very weak lines to longer wavelengths.} shows a less dense absorption line sequence. Figure~\ref{fig:01} demonstrates the appearance of the A-band at two spectral resolutions of $R=\,\frac{\lambda}{\Delta_\lambda}\,=\,104\,000$ and $R\,=\,970$. 
High signal-to-noise observations of Sirius with the Ultraviolet and Visual Echelle Spectrograph \citep[UVES,][]{Dekkeretal00} and of Feige\,110 with the FOcal Reducer and low dispersion Spectrograph 2 \citep[FORS2,][]{Appenzelleretal98}, both mounted on the Very Large Telescope (VLT) of the European Southern Observatory (ESO), are shown, as extracted from the ESO archive\footnote{\url{https://archive.eso.org}}. The Sirius spectrum was taken on January 23$^{\rm rd}$, 2002 00:27 UT at airmass 1.314 and the Feige \,110 spectrum on October 22$^{\rm nd}$, 2012 UT 00:05 at airmass 1.200.
Both stars do not show intrinsic stellar lines in the A-band regime, which allows us to study the pure effects of the molecular oxygen absorption.  At low resolution the spectrograph resolution gives the impression that potential spectral features of a target at wavelengths below $759.60$\,nm are affected by absorption. As shown in the high resolution observation this is not the case. Moreover, the low resolution spectrum gives the impression that there is no significant absorption for $\lambda>767.0$\,nm on. But in fact absorptions as strong as 80\% residual flux still show up in the range $767.0<\lambda<771\,$nm at some wavelengths.

We applied molecfit \citep{Smette2015,Kausch2015} to fit the A-band spectra by using only the full width half maximum (FWHM) of a Gaussian line spread function of the spectrograph as free parameter and not varying the Earth's atmospheric molecular profile. This profile is generated automatically by molecfit with date, time and target position (i.e. airmass) information of the observations from the file headers, based on a standard profile refined by data from the Global Data Assimilation System (GDAS)\footnote{\url{https://www.ready.noaa.gov/gdas1.php}} and the local ESO meteorological station. An excellent match is obtained, as visualized in Fig.~\ref{fig:01}. The standard deviation of the residuals of the telluric line removal amounts typically to less than about 1\%. Note that the most pronounced larger-scale residuals in the case of Sirius stem from imperfections in the data reduction -- like from Echelle order merging -- and not from the telluric correction.

In order to investigate the effect of the radial object velocity $v$ and the spectral resolving power on the telluric correction in the Fraunhofer A-band wavelength regime, we use simulated data of the Earth's atmosphere transmission $T(\lambda,a)$ in high resolution, and various target spectra $I_{\rm target}(\lambda,v)$. 
We chose a spectral resolving power of $R=250\,000$ and label the target spectrum accordingly $I_{\rm 250\,000}(\lambda,v)$ (and the resolution-dependent transmission function $T_{R}(\lambda,a)$). At this high spectral resolution no significant change of the natural line width and shape by the instrumental resolution is expected, i.e. these spectra can be used as references in the following. In particular, we assume that the quality of the telluric correction can hardly be improved incorporating higher resolved spectra.
We calculated $T_{R}(\lambda,a)$ by means of the ESO Sky Model Calculator \citep[SkyCalc,][]{skymodel_1}\footnote{online version: \url{https://www.eso.org/observing/etc/skycalc}}. This provides a comprehensive model of the sky background consisting of several components including the absorption of the Earth's atmosphere (which is computed using the same approach as employed in molecfit). 
We used the Munich Image Data Analysis System (MIDAS) package \citep{MIDAS_1} to convolve these spectra with a Gaussian kernel in order to simulate various spectrographs. 
As target spectrum $I_{250\,000}(\lambda,v)$ we simulated an exo-Earth with the same high resolution by using the Planetary Spectrum Generator \citep[PSG,][]{PSG} in its standard setup. We tested a model-based correction approach on this exo-Earth, as well with real data-supported model simulations of the post-nova V4332\,Sgr taken from the ESO data archive, own data obtained 2002 on the ESO New Technology Telescope (NTT) with the ESO Multi-Mode Instrument (EMMI) and additional information from the literature (for details see Sect.\,\ref{subsec:method_potassium}).

Since the radial velocity 
induced wavelength shift only affects the target spectrum, but not the telluric absorption features, we assume the following relation for the calculation of the simulated observed spectrum:
\begin{equation}\label{equ_3}
    F_{R}^\mathrm{obs}(\lambda,v) = T_{R}(\lambda) \,\, I_{R}(\lambda,v)\,,
\end{equation}
being $I_{R}(\lambda,v)$ the wavelength-shifted target spectrum as observable above the Earth's atmosphere with the spectral resolving power $R$.
The telluric correction function $f_R$ for a target subsequently was derived by integrating over a wavelength range $\lambda_1...\lambda_2$ of the input spectrum and the simulated observed spectrum 
\begin{equation}
\label{equ_4}
f_{R}({\lambda},v)= \frac{\int_{\lambda_1}^{\lambda_2} I_{R}(\hat\lambda,v)\,{\rm d}\hat\lambda}{\int_{\lambda_1}^{\lambda_2} F^{\rm obs}_{R}(\hat\lambda,v)\,{\rm d}\hat\lambda}\,.
\end{equation}

The integration boundaries can be chosen in principle arbitrarily. However, to compare later with observations, $\lambda_1$ and $\lambda_2$ have to correspond to individual pixel boundaries on the detector, i.e. a spectral bin.
In order to quantify the effect of the spectral resolution on the correction of the spectral feature, we introduce a binned function $\epsilon_{R}(\lambda,v)$ 
\begin{equation}\label{equ_5}
\epsilon_{R}(\lambda,v) = \frac{f_{R}({\lambda},v)}{f_{\rm 250\,000}({\lambda},v)}\,,
\end{equation}
with the $R=250\,000$ high-resolution data serving as reference where we integrate from pixel boundary to pixel boundary of the spectrograph detector chip.

\section{Applications and Results}\label{sec:applications}

\subsection{O$_2$ in direct observations of an exo-Earth} 
\label{sec:method_O2}

In the first case scenario we investigate the effect of the radial object velocity and the  spectral resolving power on the telluric correction of the O$_2$ A-band as obtainable from direct low-resolution spectroscopy of the reflected light from an exo-Earth. 
This may become feasible for an exo-Earth potentially found around a nearby star using an orbiting starshade + a ground-based ELT \citep{Mather2019}. 
We therefore concentrate on low-resolution spectroscopy in the range $R=150$ to 900. 

\begin{figure}[t!]
\medskip
    \centering
    \includegraphics[width=85mm]{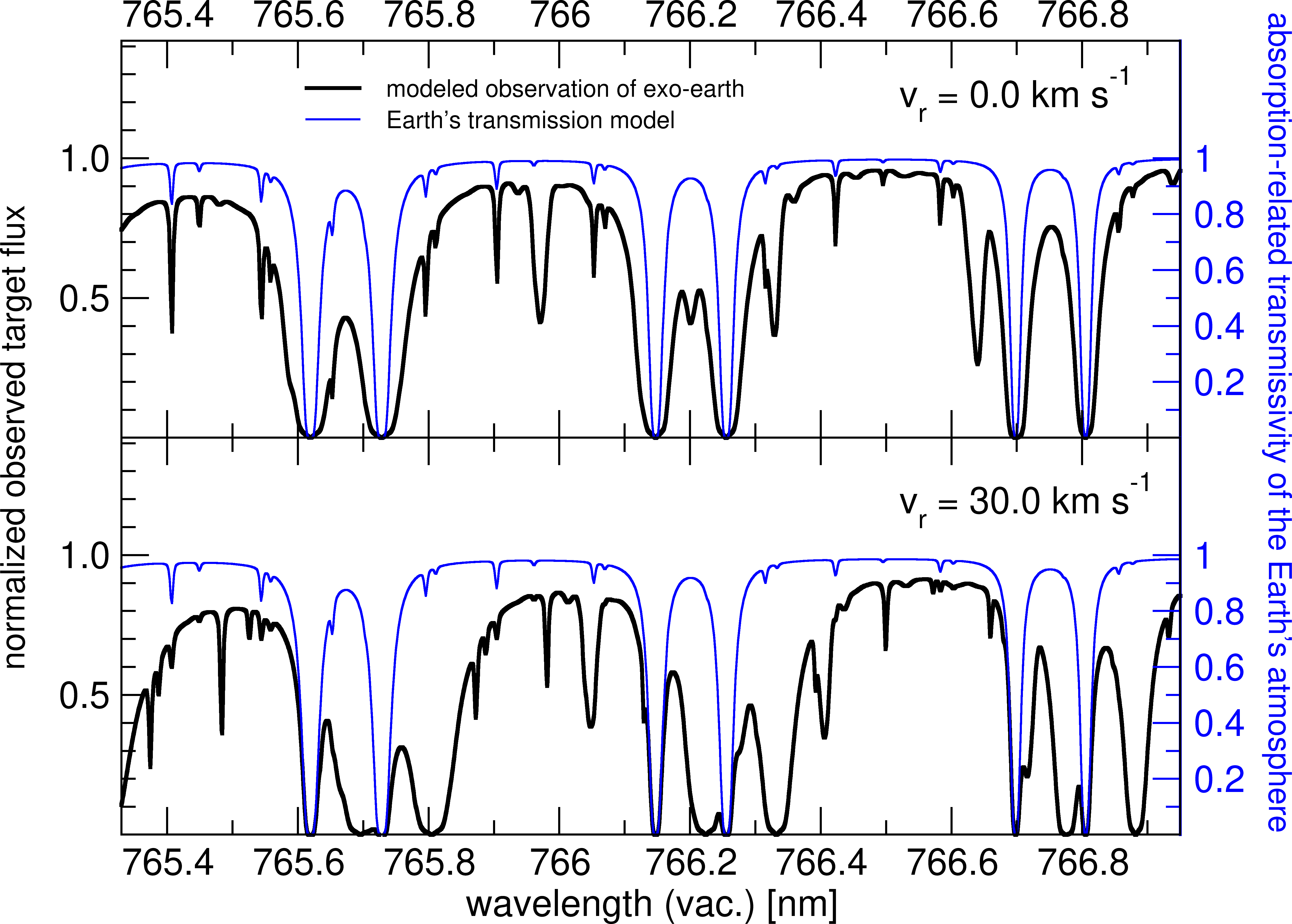}
    \caption{High-resolution model of an exo-Earth spectrum. A zoom to a subsection of the A-band at rest velocity (upper panel) and at a redshift of 30\,km\,s$^{-1}$ (lower panel) as seen from a ground based observatory is shown. The flux is normalized to the continuum outside the A-band. The applied molecular absorption model of our home planet is shown in blue overlay.}
    \label{fig:02}
\end{figure}

For that purpose we simulate an observation near zenith of an exo-Earth's atmosphere with the PSG assuming an earth-like planet in a distance of 1\,AU around a solar-type star. For convenience, we also assume a face-on orbital orientation of the system, leading to constant brightness of a half-illuminated planet during its entire orbit, and a constant $90^\circ$ scattering angle of the reflected light. In addition, we assumed no cloud coverage, incorporated an Earth's atmosphere molecular and pressure structure and assumed an average ground albedo. This is the default setup of the PSG for an earth-like planet. Although PSG offers more sophisticated models \citep[see e.g.][]{Smith2020} we focus on this relatively simple case since this is well suited to demonstrate the effects of the telluric correction.

The upper panel of Fig.\,\ref{fig:02} shows the resulting exo-Earth spectrum (for a system with $v=0.0$\,km\,s$^{-1}$) 
as seen with ground-based facilities at high spectral-resolution. The flux is normalized to the continuum outside the A-band. The applied
transmission of the Earth's atmosphere related to molecular absorption is shown as overlay. Additional spectral features in the exo-Earth
spectrum stem from the solar-type host star that illuminates this planet. Most notable is that the exo-Earth spectrum shows significantly wider absorption lines than the telluric spectrum. This is a consequence of the integration of many individual light paths of different lengths in the exo-atmosphere due to  scattering and reflection on the entire surface, whereas the light rays from the exo-Earth just pass from one direction the Earth's atmosphere until they reach the telescope.
Furthermore, multiple scattering near the limb leads to wide shallow wings in the PSG model which nearly turn into a quasi continuum absorption. The PSG model reaches unity only outside the band.
The absorption features of the Earth's spectrum fall exactly in the regime where the exo-Earth does not have any measurable flux. Thus, even if features can not be resolved at low resolution spectroscopy, no telluric absorption correction should be applied at all.

The lower panel of Fig.~\ref{fig:02} shows the resulting observable exo-Earth spectrum for a system with a differential radial velocity $v=30.0$\,km\,s$^{-1}$. As the absorption features of the Earth's atmosphere stay at rest velocity, while those of the exo-Earth move, the Earth's atmosphere now absorbs a certain fraction of the target's light. The observed spectrum significantly changes as function of the radial velocity. 
We expect for the velocity range in the Galactic vicinity a quasi-periodic pattern of similar exo-Earth features every $v\approx215$\,km\,s$^{-1}$ because of the periodic line pair nature of the P-branch of the A-band. This period is derived when considering the entire P-branch, where the strongest lines with lower quantum numbers dominate the pattern. In case of use of sub-fractions of the region, significantly different values may be deduced (see Sect.~\ref{subsec:method_potassium}).
\begin{figure}[t!]
\medskip
    \centering
    \includegraphics[width=85mm]{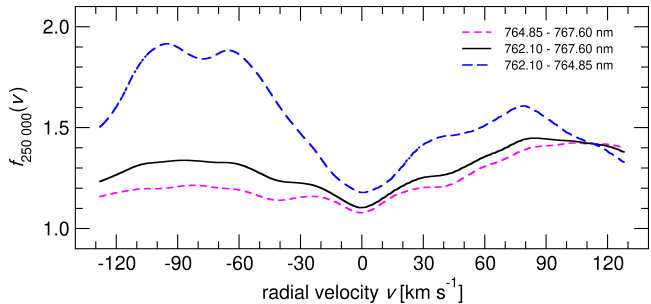}
    \caption{Telluric correction function (see Eqn.\,\ref{equ_4}) for our test case of an exo-Earth spectrum at spectral resolution $R=250\,000$ as a function of the radial velocity $v$, calculated as the ratio of the integrated fluxes above the Earth's atmosphere and the observation on the ground. Results for the integration over three wavelength ranges within the P-branch of the O$_2$ A-band are shown, as indicated.}
    \label{fig:03}
\end{figure}

As abundance effects of exoplanets can be investigated only by non-completely saturated lines, we focus the investigation of the effects  of the radial velocity  in low resolution spectroscopy  on the P-branch of the A-band. We define three wavelength ranges: (a) $\lambda=762.10...767.60\,$nm, covering all the range with more than 5\% A-band average absorption in low resolution spectroscopy (see~panel C in Fig.~\ref{fig:01}), and further split this range in two regions (b) $\lambda=762.10...764.85\,$nm, and (c) $\lambda=764.85...767.60\,$nm, defined by the region where O$_2$ lines are fully saturated in the cores and those which are not fully saturated (see~panels A and B in Fig.~\ref{fig:01}). We then integrate the flux in these ranges for both, the exo-Earth spectrum $I_{\rm 250\,000}(\lambda,v)$ as it would be observed above the Earth's atmosphere and
$F_{\rm 250\,000}^{\rm obs}(\lambda,v)$ as observed on the ground at high resolution, and calculate the corresponding flux ratio $f_{250\,000}(\lambda,v)$ (see~Eqn.~\ref{equ_4}) using the integration boundaries mentioned above. 
This gives a measure of the 
telluric absorption for integrated line strengths in the given wavelength ranges. 
The behavior of this factor is shown in Fig.\,\ref{fig:03}, calculated for the radial velocity range between $\Delta v=\pm130\,$km\,s$^{-1}$ with a grid step $\delta v=$1\,km\,s$^{-1}$. This velocity range should cover practically all cases of motions of exoplanetary systems in the solar neighborhood.
In fact, the dependency of the integral in the numerator on the velocity $v$ in Eqn.\ref{equ_4} is marginal using such large integration ranges as employed here and thus may be neglected. 

Nearly no (additional) flux is absorbed by Earth's atmosphere at zero radial velocity of the target. Thus, the telluric correction function remains close to unity in all three wavelength ranges. This changes for non-zero radial velocity, where the telluric correction factor can amount to almost a factor of 2 in the most extreme case. We have to stress here that {\em knowledge of the radial velocity of the target is essential for the telluric correction} -- which is trivial at the natural high-resolution scale discussed here, but becomes a priori difficult to be determined at low spectral resolution imposed by the measurement with a realistic instrument.
\begin{figure}[t!]
\medskip
    \centering
    \includegraphics[width=85mm]{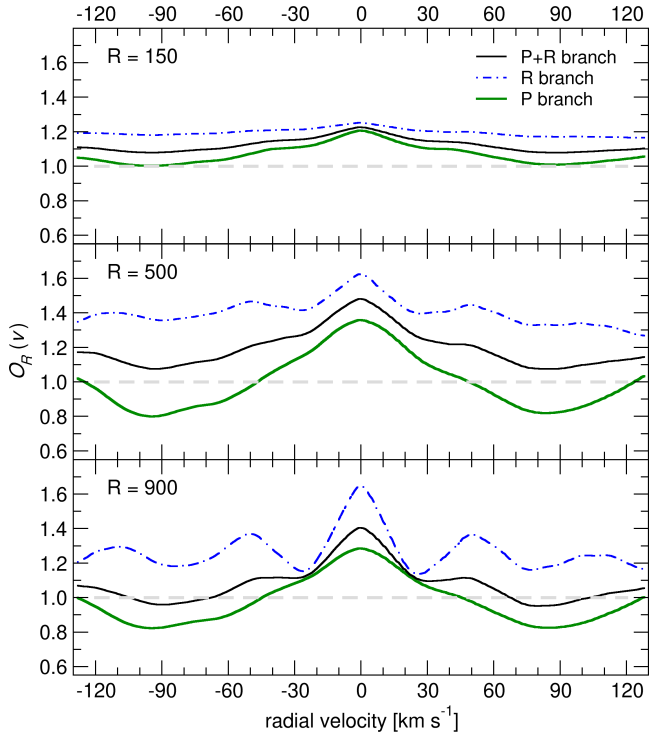}
    \caption{Ratio $O_{R}(v)$ for three different resolving powers for the P and R-branches individually and for the total A-band.}
    \label{fig:04}
\end{figure}

At low spectral resolution the entire A-band is covered by a low number of spectral bins. For investigating the quality of the telluric correction as a function of the spectral resolving power, we extend our considerations to both, the P- and the R-branch, and the entire A-band (P+R branch) using $I_{R}(\lambda,v)$ and $F_{R}^{\rm obs}(\lambda,v)$ spectra with resolving power of $R=150$, $500$, and $900$, respectively. We applied the classical telluric correction method (Eqn.\,\ref{equ_2}) to the observed spectrum $F_{R}^{\rm obs}(\lambda,v)$, hereafter denoted as $I_{R}^{\rm obs,corr}(\lambda,v)$ and calculated the integrated flux for the P-branch ($\lambda_{\rm start}=762.1\,$nm, $\lambda_{\rm end}=767.0\,$nm) the R-branch ($\lambda_{\rm start}=759.6\,$nm, $\lambda_{\rm end}=761.9\,$nm) and the entire A-band ($\lambda_{\rm start}=759.6\,$nm, $\lambda_{\rm end}=767.0\,$nm) individually for the same velocity grid as above. The same integration was applied for the full resolution spectra as it was used to derive Fig.\,\ref{fig:03}. 

The velocity-dependent ratio of observed to real integrated fluxes 
\begin{equation}\label{equ_6}
    O_{R}(v) = \frac{\int_{\lambda_{\rm start}}^{\lambda_{\rm end}}{I_{R}^{\rm obs,corr}(\lambda,v)\,{\rm d}\lambda}}{\int_{\lambda_{\rm start}}^{\lambda_{\rm end}}{
I_{\rm 250\,000}(\lambda,v)\,{\rm d}\lambda}}
\end{equation}
is linked to a sum of the bins of the function $\epsilon_{R}(\lambda,v)$ (see Eqn.\,\ref{equ_5}) over the wavelength region:
\begin{equation}\label{equ_7}
    O_{R}(v) \equiv \frac{\sum_{\lambda_{\rm start}}^{\lambda_{\rm end}}{\epsilon_{R}(\lambda,v)}}{{\sum_{\lambda_{\rm start}}^{\lambda_{\rm end}}{
\epsilon_{\rm 250\,000}(\lambda,v)}}}\,.
\end{equation}
Figure\,\ref{fig:04} shows that the application of the classical telluric correction method leads to both, over- and under-correction of the fluxes as a function of the radial velocity $v$. This effect is strongly depending on the spectral resolving power as well. At very low resolution only over-correction occurs, whereas the situation changes with increasing resolving power. The strongest error in form of an over-correction always occurs at low radial velocities (see above). In addition, the R-branch is in general more affected than the P-branch -- a result of the much denser oxygen absorption line forest in that wavelength range.

\begin{figure}
\medskip
    \centering
    \includegraphics[width=85mm]{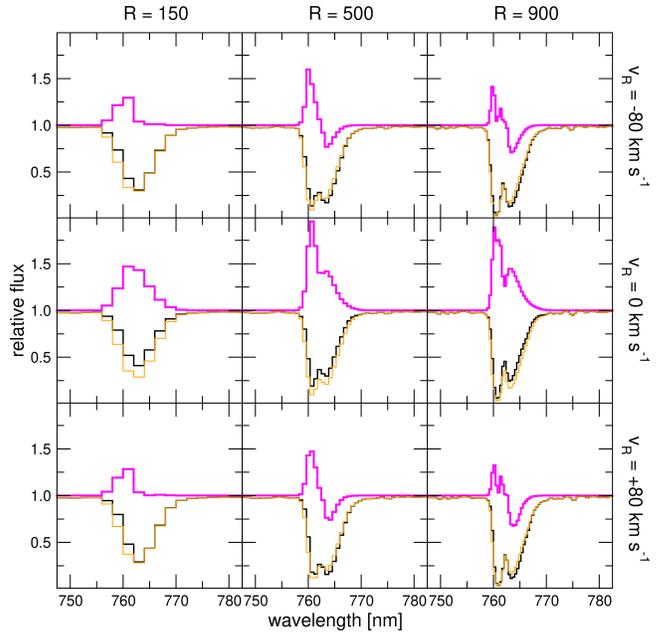}
    \caption{Wavelength-resolved spectra of an exo-Earth for different spectral resolutions and radial velocities. Displayed are the normalized input spectrum $I_{R}(\lambda,v)$ of the target (orange lines), the normalized spectrum $I_{R}^{\rm obs,corr}(\lambda,v)$ after a classical telluric correction (black lines) and the resulting function $\epsilon_{R}(\lambda,v)$ (magenta lines).}
    \label{fig:05}
\end{figure}

Figure~\ref{fig:05} compares the input spectra $I_{R}(\lambda,v)$, $I_{R}^{\rm obs,corr}(\lambda,v)$, and the resulting function $\epsilon_{R}(\lambda,v)$ (see Eqn.\,\ref{equ_5}) in the entire A-band. We show the cases of radial velocity $v= 0$\,km\,s$^{-1}$ and $v=\pm80$\,km\,s$^{-1}$ as they representing approximately the regions with large total over- and underestimation of the flux errors after applying the telluric correction (see Fig.\,\ref{fig:04}). 
Two conclusions can be drawn from this. First, the telluric correction in our test case always leads to an over-correction at very low resolution ($R=150$). However, a near-perfect restoration of the original flux in the P-branch can be obtained around $v\approx\pm80$\,km\,s$^{-1}$. Second, the R-branch is always over-corrected at the higher resolutions of 500 and 900, but the P-branch can be both over- as well as under-corrected. A 'sweet spot' of near-perfect correction of the diagnostic P-branch, i.e. for $O_{R}(v)\approx1$ (see Fig.\,\ref{fig:04}), is reached around radial velocities of $\pm$45\,km\,s$^{-1}$. If the orbits of the exo-Earth and therefore its radial velocity would be known e.g. from previous imaging at different epochs, one could aim to schedule the spectroscopic observations such that the telluric impact would be minimized. On the other hand, knowledge of the  function $\epsilon_{R}(\lambda,v)$ could be used to recover the original exo-Earth flux as seen above the atmosphere in a second step after the classical telluric correction.
Also CO$_2$ and methane bands show such regular patterns in the near infrared regime, and therefore also lead to distinct {\em sweet spots}, each at individual velocities. The situation for molecules like water vapor is different since their {\em sweet spots} are less distinct due to their more complex band structure.

\subsection{Potassium \ion{K}{1}\,(D$_{1,2}$) lines at 766.5 and 769.9\,nm} \label{subsec:method_potassium}

In the second case scenario we investigate the effect of the radial target velocity and the spectral resolving power on the telluric correction of existing observations. We address the case of target emission lines falling in the wavelength range of the Fraunhofer O$_2$ A-band.

To date, the most prominent target showing the potassium \ion{K}{1}\,(D$_{1,2}$) lines in emission is the post nova 
V4332\,\,Sgr \citep{Banerjee2004,Kimeswenger06}. The low excitation of 1.61\,eV 
makes them a very good indicator for warm but not ionized nebulae or for dense circumstellar 
shells \citep{Hobbs1974, Welty2001}. Although the cosmic abundance of potassium is about 15 times lower than that 
of sodium \citep{StandardAbundance},  
the significantly lower excitation energy, compared to the 2.10\,eV for 
the well studied \ion{Na}{1}\,(D$_{1,2}$) lines at 589.00 and 589.59\,nm, 
makes the potassium lines still a better tracer by higher population of the upper level for gas at temperatures below 850\,K. 
However, since their wavelengths are at  \ion{K}{1}(D$_2$)\,=\,766.4899\,nm and \ion{K}{1}(D$_1$)\,=\,769.8964\,nm, respectively, they are located in the P-branch of the telluric O$_2$ A-band.  
Moreover, these lines are also present in the airglow spectrum arising from the Earth's atmosphere, requiring a decent sky subtraction, and are therefore rarely used. Due to the asymmetry of the A-band, the \ion{K}{1}(D$_2$) line is more affected by the O$_2$ absorption. In the rest frame the \ion{K}{1}(D$_2$) coincides with one of the O$_2$ lines whereas the \ion{K}{1}(D$_1$) is unaffected \citep{Noll2019}. Since their line ratio is essential to derive physical parameters like density and optical depth of the target object it is crucial to have knowledge on its radial velocity to ensure an appropriate application of the telluric correction. 

In a first step to investigate the telluric correction effects we simulate the same assumptions as \citet{Banerjee2004} used in their observations obtained at the high altitude station Himalayan Chandra Telescope (HCT) of the Indian Astronomical Observatory \citep{HCT} using the Hanle Faint Object Spectrograph Camera (HFOSC) for the modeling of the \ion{K}{1}(D$_{1,2}$) lines in V4332\,\,Sgr: a Gaussian line profile was adopted with the FWHM ranging from 0.05\,nm to 0.40\,nm. This corresponds to a target intrinsic turbulent, rotational or expansion velocity of about 20 to 155\,km\,s$^{-1}$ at a spectral resolving power of $R \approx 900$. In case of nebulae and circumstellar material the FWHM is limited to $\,\gtrapprox\,$0.1\,nm  due to their internal physics of local turbulence, temperature broadening and large scale motions like rotation and expansions. We therefore focus on those cases. The lowest values will be rare (e.g. weak nightglow emission in geophysical investigations like in \citet{Noll2019}). Therefore they are shown here as well to demonstrate that these will even give higher complexity then. For certain velocities extremely narrow lines limit the use of the methods here (see below).
Although the time of observations was not published in \citet{Banerjee2004}, we are able to estimate the airmass because the target was only observable about 30 minutes at the beginning of that night after astronomical twilight ended (see Table\,\ref{obsV4332}).
We sampled the radial velocity range $v\,\in[\pm190]$\,km\,s$^{-1}$, chosen to represent typical values for Galactic targets. 

\begin{figure}
\medskip
    \centering
    \includegraphics[width=85mm]{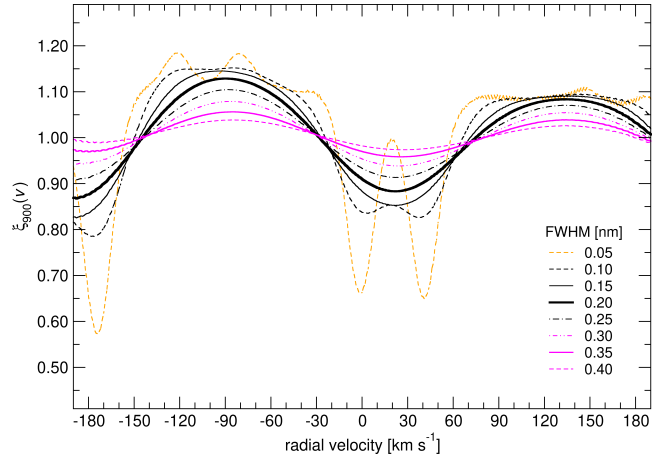}
    \caption{Quotient $\xi_{900}(v)$ (see Eqn.\,\ref{equ_8}) of the \ion{K}{1}\,(D$_{1,2}$)  line ratios after a classical telluric correction of the A-band as a function of target radial velocity. The different line styles parameterize the FWHM of the \ion{K}{1}\,(D$_{1,2}$) lines (in nm), as indicated.}
    \label{fig:06}
\end{figure}
In order to determine the effect of the telluric correction on the line flux ratio arising from the low resolution, we calculated the line-ratio quotient for classically corrected spectra at low and high resolving power {$R = 900$ and $R = 250\,000$}. While the spectrum can be corrected classically with very high accuracy at $R = 250\,000$ (see Eqn.\,\ref{equ_2}) because the telluric lines are well resolved, the low resolution data are affected by the induced errors. The line flux ratios \ion{K}{1}\,(D$_{1,2}$) $I$(766.5\,nm)/ $I$(769.9\,nm) were determined for both cases individually, and subsequently their ratio was calculated for the entire $v$ range.
We thus define this quotient of the line ratios after telluric correction $\xi_R(v)$ as
\begin{equation}\label{equ_8}
     \xi_R(v) = \frac{\eta_R(v)}{\eta_{250\,000}(v)}
\end{equation}
where the line ratio after the telluric correction $\eta_R(v)$ takes the form
\begin{equation}
    \eta_R(v)=\frac{\int_{\rm K I(D2)} I_{R}(\lambda,v) {\rm d}\lambda}{\int_{\rm K I (D1)} I_{R}(\lambda,v) {\rm d}\lambda}\,\,.
\end{equation}

This is equivalent to deriving $O_{R}(v)$ (see Eqn.\,\ref{equ_6}) in two small regions $\lambda_{\rm start}^{\rm D1}...\lambda_{\rm end}^{\rm D1}$ and $\lambda_{\rm start}^{\rm D2}...\lambda_{\rm end}^{\rm D2}$ around each of the two lines and build the quotient of these two values. 
Like in the case of the exo-Earth oxygen band, both, over- and underestimates are realized, see Fig.\,\ref{fig:06}. Notable is the strong dependency of the line-ratio quotient on the FWHM of the target lines. Once the line FWHM is lower than 0.01\,nm (which corresponds to  $3.9\,$km\,s$^{-1}$) the calculation of the line-ratio quotient becomes mathematically unstable due to divisions by very small numbers in Eqn.\,\ref{equ_4} and also physically meaningless due to missing observed line flux contributions if the blue line of the pair at $\lambda_0=766.4899\,$nm coincides exactly with a fully saturated O$_2$ line. This occurs at radial velocities of $-$2.3\,km\,s$^{-1}$ or +40.7\,km\,s$^{-1}$), respectively, and is repeated by shifts of 245\,km\,s$^{-1}$ due to the quasi-periodic pattern. The somewhat higher value than in the previous section for the period originates from the fact that only the weaker red part of the P-branch is involved here. The velocity shifts between the line pairs steadily grow a bit towards higher quantum numbers (see~Fig.\,\ref{fig:01} panel A).

\begin{figure}[t!]
\medskip
    \centering
    \includegraphics[width=85mm]{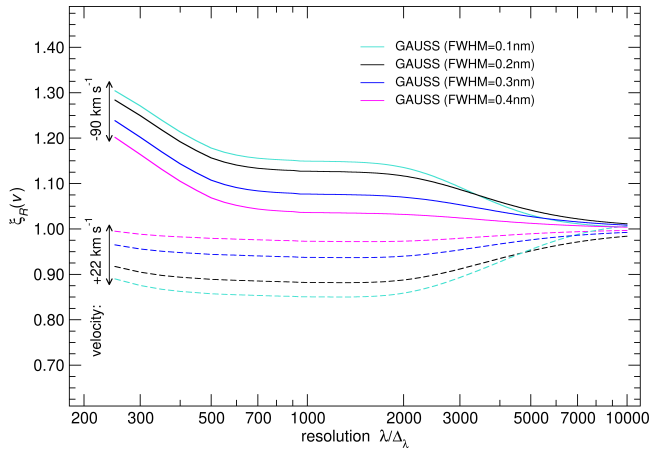}
    \caption{Effect of the spectroscopic resolution for the two velocities that give the strongest over- and underestimate of the \ion{K}{1} line ratio after a classical telluric correction. The different line styles parameterize the FWHM of the \ion{K}{1} lines, as indicated.}
    \label{fig:07}
\end{figure}

For the investigation on the spectral resolving power influence we focus on the extreme cases at $v=-90$ and +22\,km\,s$^{-1}$  for most astrophysical objects with FWHM$\,>\,$0.1\,nm (see Fig.~\ref{fig:06}). Figure~\ref{fig:07} shows the line-ratio quotient as function of the spectral resolving power for these two $v$ cases and for four different values of the FWHM. The error increases nearly linearly towards decreasing resolutions below $R < 500$, reaching values of up to 30\%. Between $500 < R < 3000$ the resolving power has nearly no effect on the quotient, whereas for higher resolutions the discrepancy slowly vanishes. With spectral resolutions of $R \gtrsim 10\,000$ the oxygen line pairs are fully resolved, which is a pre-condition for the classical telluric correction to work properly without target model based corrections.

\begin{figure}
\medskip
    \centering
    \includegraphics[width=85mm]{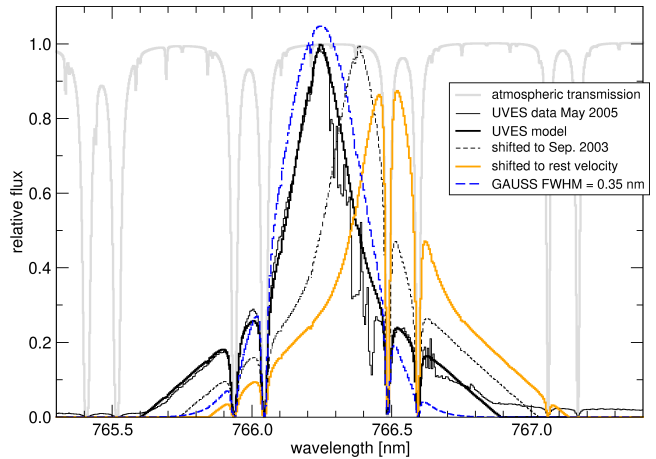}
    \caption{The observed ESO UVES spectrum of the \ion{K}{1}(D$_2$) line of V4332\,Sgr. Displayed are also the adopted line-profile model, the model after a shift to the radial velocity of September 2003 (the epoch of the HCT HFOSC observations) and at laboratory rest velocity. A Gaussian with FWHM\,=\,0.35\,nm multiplied with the telluric transmission , as used in the analysis of the Subaru HDS observation \citep{Subaru} is shown as well. The applied atmospheric absorption at the airmass of the UVES observation without the continuum extinction by scattering of about 4\% in that red wavelength range is shown as overlay. For easier comparability the hypothetical observations at other velocities induced by the different observational epochs were simulated with the same near-zenith observations like obtained with UVES. The peak of the original UVES observation is used to normalize the flux. All model data are intentionally sampled in step sizes of the UVES detector chip.}
    \label{fig:08}
\end{figure}

After the model comparisons we address the actual observations of V4332\,\,Sgr. Table~\ref{obsV4332} gives an overview on these observations including information on the employed instrument and instrument configuration, observing date and time, airmass, the achieved spectral resolving power and a set of velocities at the time of the observations: the barycentric correction $v_\mathrm{bc}$, the UVES observation velocity $v_\mathrm{obs}$ and calculated observational velocities for the other epochs $v^\mathrm{calc}_\mathrm{obs}$.

We used the UVES data with the best resolution to derive an observed velocity $v_{\rm obs} = -96$\,km\,s$^{-1}$ which defines a barycentric system velocity of $-$72.9\,km\,s$^{-1}$. 
Only \citet{Subaru} reported previously a velocity measurement. They found a heliocentric velocity of $-69.0\pm2.9\,{\rm km}\,{\rm s}^{-1}$, 
which corresponds well with our value.
To calculate the barycentric corrections for the various observations the \mbox{MIDAS} system command \mbox{COMPT/BARY} was used.
The effect of Earth's orbital motion on the \ion{K}{1}(D$_2$) line of V4332\,Sgr is clearly seen against the (static) telluric background lines (see Fig.\,\ref{fig:08}).  While in the UVES observations the \ion{K}{1}(D$_2$) line core is not affected by the telluric O$_2$ absorption, it was partly weakened in the HCT HFOSC data by  \citet{Banerjee2004}.

The high-resolution UVES data allows for the first time the intrinsic line shape to be determined in detail. The best model was found by incorporating a Gaussian main core with a FWHM of 0.20\,nm, and a wider wing, shaped like a triangle. However, the UVES observations were overexposed in the line center, thus they reach the nonlinear sensitivity regime of the instrument detector. Therefore, a full calibration is unfortunately not possible.

\begin{figure}
\medskip
    \centering
    \includegraphics[width=85mm]{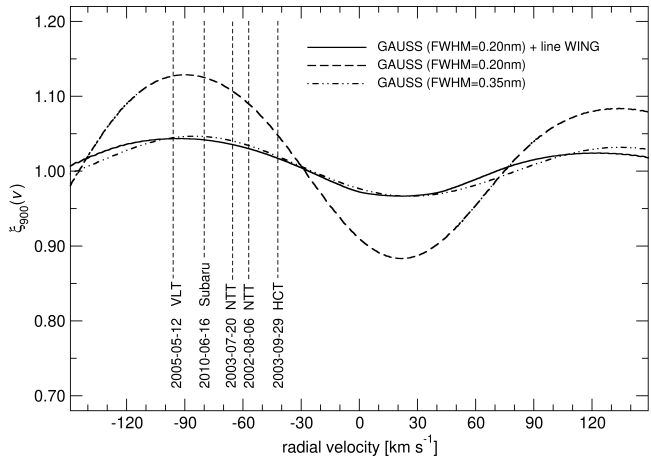}
    \caption{Quotient $\xi_{900}(v)$ of the \ion{K}{1}\,(D$_{1,2}$) line ratios after a classical telluric correction, considering also a line shape as derived from the UVES data. Radial velocities at the times of the available observations of V4332\,Sgr are indicated.}
    \label{R900_full}
\end{figure}

However, the UVES data allow to investigate the influence of the line shape on the line-ratio quotient introduced by a classical telluric correction. Three cases are compared, the intrinsic profile with a Gaussian core of FWHM of 0.2\,nm plus broad wings, only the Gaussian profile with FWHM of 0.2\,nm and another Gaussian model with FWHM of 0.35\,nm, which was found to be the best fit to intermediate resolution spectra taken in 2010 with the High Dispersion Spectrograph (HDS) spectrograph at the Subaru telescope \citep{Subaru}. We normalize this model to the same flux of our broader-wing Gaussian model leading to a slightly higher peak intensity. The resulting line-ratio quotient is compared in Fig.~\ref{R900_full}.  
While the broader-wing model shows a significantly smaller error than the pure Gaussian with the same FWHM, rather close agreement of the deviation of our broader-wing Gaussian and the Subaru model is achieved. 
As a summary, we conclude that the line shape is a crucial quantity having a significant influence on the telluric correction quality. It should be emphasized that the case at hand shows that two of the three very different line shapes happen to produce nearly identical corrections. The solution for reversing the process is therefore ambiguous.

Finally, one may ask how large an effect the refined telluric correction has on the physical interpretation of the observations. The line ratio of the potassium lines was employed by 
\citet{Banerjee2004} to study the optical depth $\tau$ of the emitting plasma, yielding $\tau\sim4.5$.
By virtue of a favorable radial velocity of the target at the epoch of their spectroscopy (see Fig.~\ref{R900_full}) the classical telluric correction resulted in an error of the potassium line ratio of a mere 3\%.
The resulting correction of the optical depth is thus about $ \Delta\tau \approx +1.0$. 
\relax

\begin{table*}[t]
\caption{Observations of V4332\,Sgr (in chronological order): instrument, instrument configuration, date, time, airmass, reso\-lution, barycentric corrections $v_{\rm bc}$, UVES observation velocity $v_{\rm obs}$ and calculated observational velocities for other~epochs~$v_\mathrm{obs}^{\rm calc}$.}\label{obsV4332}
\noindent\hspace{-19mm}{\begin{tabular}{l l c c c c r r r}
\tableline
instr. & instrument configuration & date & time & airmass &  $R$ & $v_{\rm bc}$ & $v_{\rm obs}$ & $v_{\rm obs}^{\rm calc}$ \\
& & & UT & &  $\frac{\lambda}{\Delta_\lambda}$ & [km s$^{-1}$] &[km s$^{-1}$] & [km s$^{-1}$] \\
\tableline
\tableline
EMMI\tablenotemark{\small\rm 1}  & Grism\#2 380-920\,nm, slit: 1\farcs0 & 2002/08/06 & 00:19 & 1.17 & $\approx$800 & $-$15.4  & & $-$58\phantom{\small a}\\
 & Grism\#2 380-920\,nm, slit: 1\farcs0 & 2003/07/23 & 00:43 & 1.21 & $\approx$800 & $-$7.6  & & $-$65\phantom{\small a} \\
HFOSC\tablenotemark{\small\rm 2} & Grism, slit: 1\farcs3 & 2003/09/29 & 13:30-15:00 & 1.60-1.95 & $\approx$900 &$-$29.7 & & $-$43\phantom{\small a} \\
UVES\tablenotemark{\small\rm 3} & RED860 665-1023 nm, slit: 1\farcs0 & 2005/05/12 & 06:43 & 1.07 & 42\,300 & $+$23.0 & $-$96\tablenotemark{\small\rm a} &  \\
HDS\tablenotemark{\small\rm 4} & SC46 680-798 nm, slit: 0\farcs5 & 2009/06/16 & 12:32 & 1.38 & 21\,000 &  \phantom{l}$+$8.2 & & $-$81\phantom{\small a}  \\ 
\tableline
\end{tabular}}

\noindent{\small
{a) velocity measured here in this work and used as 
reference for the system velocity of $-$72.9\,km\,s$^{-1}$ }\protect{\newline}
{1) \citet{Kimeswenger06}}\protect{\newline}
{2) \citet{Banerjee2004}}; UT estimated from object visibility starting during twilight already.\protect{\newline}
{3) this work}\protect{\newline}
{4) \citet{Subaru}}}
\end{table*}

\relax

However, because of the highly non-linear dependency of the optical depth, the result is very sensitive to larger corrections. For example a reduction of the line ratio by 10\% would result already in the solution $ \tau \rightarrow \infty $.

\section{Conclusions} \label{sec:conclusions}
The quantitative analysis of spectral features in wavelength regimes highly affected by deep and variable Earth's atmosphere absorption is  a difficult matter. In this work we revealed that the application of the classic telluric correction (Eqn.\,\ref{equ_2}) on spectra taken with low resolving powers ($R\lesssim10\,000$) induces shortcomings in the recovery of target features. Spectral properties, e.g. the line shape and the spectral resolution, as well as the radial velocity of the target, lead to errors of up to a factor of 2 in the correction in our two test cases based on the oxygen O$_2$ A-band and the \ion{K}{1}\,(D) doublet. We also demonstrated that only the usage of high-resolution spectra for both, the target and the telluric transmission spectrum leads to a reliable line recovery. We therefore conclude that a high spectral resolution model of the target spectrum, as well as a high resolution transmission curve of the Earth's atmosphere are essential to derive the required function to correct the observed spectrum for these effects even if low-resolution spectra are investigated. Note that completely saturated lines might occur in telluric absorption bands. Since the telluric correction incorporates the division by $T(\lambda)$ (Eqn.\,\ref{equ_2}), the corrected flux reaches infinity, which limits the applicability of this method in general. However, in low resolution spectra, this effect of fully saturated absorption is not obvious (see Fig.\,\ref{fig:01}) and is therefore often not considered.

The method applied to the two test cases can be generalized to any wavelength region with strong spectral features in the real, but in the instrument unresolved, target spectrum $I_{\rm 250\,000}(\lambda,v)$ and  real transmission curve $T(\lambda,a)$, e.g.~for water vapor observations in brown dwarfs and nearly all types of planets (in the solar system as well as in exoplanetary systems), molecular features in Herbig-Haro objects, near infrared H-band spectra of CO, CO$_2$ or CH$_4$ in late-type stars, or to highly-redshifted Lyman-$\alpha$ emitters at redshifts $z > 5.2$ in cosmological investigations. The function $\epsilon_{R}(\lambda,v)$, and the resulting correction function, then strongly depend on the target case.  To calculate the required high-resolution transmission $T(\lambda,a)$ for applying our method comprehensive numerical codes, e.g. the Line-By-Line Radiative Transfer Model \citep[LBLRTM,][]{LBLRTM} are available and widely in use \citep{Seifahrtetal10,skymodel_1,Bertaux2014,Gullikson2014,Smette2015,Kausch2015}.

Adopting a high-resolution target model spectrum is the more crucial part -- e.g., future facilities will barely facilitate low-resolution spectra of exoplanets at decent signal-to-noise ratio to be obtained. One would need to adopt grids of high-resolution model spectra for the target (large model grids of exoplanet spectra from various groups are currently becoming available), perform the refined telluric correction at high resolution and then compare the results with the low-resolution observations to find the best-matching model. Another possibility is an iterative process based on a classically corrected spectrum as first guess and subsequent refining and constraining of the incorporated parameters. In particular, the determination of the target's radial velocity $v$ is a crucial parameter, which can be derived by means of spectral features in the target spectrum outside the strong telluric absorption or by direct imaging of the orbit of an exoplanet. 
As the radial velocity component of Earth's orbital motion produces a peak-to-peak value of $\Delta v^\mathrm{max} = 60 \times \cos\beta~{\rm km}\,\,{\rm s}^{-1}$, with $\beta$ being the ecliptic latitude of the target, one may aim at timing the observations for crucial targets. However, different molecular features in the target spectrum may require different velocity offsets in order to minimize the
impact of remnant systematic errors after the telluric correction, thus requiring repeated observations.

\acknowledgements{
W.~Kausch was partly funded by the Hochschulraumstrukturmittel provided by the Austrian Federal Ministry of Education, Science and Research (BM:BWF). This research has made use of the SIMBAD database, operated at CDS, Strasbourg, France and of the NASA's Astrophysics Data System. The investigation uses data from the ESO data archive \url{http://archive.eso.org} from Obs IDs 069.D-0486, 071.D-0084, 075.D-0511, 090.D-0081 and 266.D-5655. We thank the anonymous referee for constructive criticism that helped to improve the manuscript.}

\hspace{2cm}

\facilities{ESO VLT:Antu (FORS2), ESO VLT:Kueyen (UVES), ESO NTT (EMMI)}
\software{molecfit \citep{Smette2015,Kausch2015}, SkyCalc \citep{skymodel_1}, MIDAS \citep{MIDAS_1,MIDAS_2}, PSG \citep{PSG}, LBLRTM \citep{LBLRTM}}

\newpage
\bibliographystyle{aasjournal}
\bibliography{references}{}

\end{document}